# Comb-enabled spectral-domain image transport through perturbation-prone multimode fibers

Maohan Li[1#], Zijian Wang[1,2#], Bowen Sun[1], Zhuoren Wan[1], Xiangze Ma[1], Xiuxiu Zhang[1], Yuan Chen[1], Mei Yang[1], Qi Wen[1], Zhaoyang Wen[1,2], Ming Yan[1,2]*, and Heping Zeng[1,2,3,4]

[1]State Key Laboratory of Precision Spectroscopy, and Hainan Institute, East China Normal University, Shanghai, China

[2]Chongqing Key Laboratory of Precision Optics, Chongqing Institute of East China Normal University, Chongqing 401120, China

[3]Jinan Institute of Quantum Technology, Jinan, Shandong 250101, China

[4]Shanghai Research Center for Quantum Sciences, Shanghai 201215, China

#These authors contributed equally.

* Corresponding author: *myan@lps.ecnu.edu.cn*

**Abstract:**

Multimode fibers (MMFs) offer a compact platform for imaging, sensing, and information transport, but their practical deployment is hindered by sensitivity to fiber perturbations, which alter modal coupling and invalidate conventional speckle-based calibrations. Here, we demonstrate perturbation-resilient image transport through MMFs by combining image-to-spectrum encoding with dual-comb spectroscopy. Two-dimensional images are converted into comb-line-resolved spectral signatures before fiber transmission, allowing spatial information to be carried in the spectral domain rather than in the output speckle field. After propagation, dual-comb heterodyne detection maps the encoded spectrum into the radio-frequency domain, enabling massively parallel spectral readout with a single photodetector. Neural-network-assisted compressive reconstruction further enables high-fidelity imaging from sparse, noisy, and spectrally aliased measurements. Our approach achieves Pearson correlation coefficients exceeding 0.9 under strong fiber perturbations and supports frame rates up to 2.5 MHz, allowing the observation of transient switching dynamics in a digital micromirror device. These results establish a powerful tool for robust, real-time image transport through flexible MMFs, with potential applications in remote sensing and fiber-based optical instrumentation.

## Introduction

High-throughput image transport through fiber-scale probes is important for applications ranging from endoscopy [1] and remote inspection [2] to optical communications [3]. Multimode fibers (MMFs) are particularly attractive because their large modal capacity enables high-dimensional spatial information transfer within a compact geometry [4]. Their large core diameters and high numerical apertures further provide efficient light collection and delivery with relaxed alignment tolerances compared with single-mode fibers [5, 6]. However, reliable imaging through MMFs remains challenging because fiber bending, twisting, and mechanical vibrations induce complex modal coupling, generating rapidly varying speckle patterns that disrupt image transport [7, 8].

Early approaches relied on transmission-matrix characterization under static conditions, but their performance degrades rapidly as the fiber state changes [8]. Although fast matrix acquisition [9, 10] and adaptive wavefront shaping [11] can restore focusing or partial field control, full transmission-matrix reconstruction becomes increasingly impractical as the number of guided modes grows [9]. Alternative methods, including memory-effect imaging [12], guide-star techniques [13], and external sensing [14], reduce calibration requirements but introduce constraints on geometry, accessibility, feedback, or system complexity. More recently, deep-learning approaches [15-18] have improved robustness through multi-state training and spatio-spectral control, yet they require extensive training data and remain limited to perturbations represented in the training set. Consequently, robust, high-speed imaging through MMFs under unknown and time-varying perturbations remains a major challenge.

Instead of compensating for perturbation-induced changes in the spatial field, a fundamentally different strategy is to encode images into a domain that is intrinsically robust to modal perturbations. In long MMFs, spectral responses can exhibit remarkable stability and transferability despite substantial variations in modal distributions, suggesting that spatial information may be transported through the spectral rather than spatial domain. Although nonlinear image-to-spectrum encoding has recently demonstrated this possibility [19], it relies on nonlinear frequency conversion and camera-based spectrometric detection, which constrain efficiency, speed, and scalability.

Optical frequency combs, with their large number of precisely defined spectral modes, provide a powerful framework for high-speed spectral encoding and detection [20–22]. Among these techniques, dual-comb spectroscopy (DCS) achieves scan-free acquisition by mapping optical spectra into radio-frequency (RF) beat signals through the interference of two combs with slightly offset repetition rates, enabling massively parallel readout using a single photodetector [23–25]. This approach has transformed fields ranging from molecular spectroscopy [23–30] and remote sensing [31, 32] to optical ranging and imaging [33–36]. While DCS-assisted MMF imaging has recently been demonstrated [37], prior implementations remain based on speckle reconstruction and have not explored spectral-domain image transport as a route to perturbation resilience.

In this paper, we introduce a comb-enabled approach for spectral-domain image transport through perturbation-prone MMFs. Spatial information is encoded into comb-line-resolved spectral signatures that remain robust to fiber perturbations and are retrieved in parallel through dual-comb detection. A neural-network decoder reconstructs images directly from the encoded spectra and enables compressive recovery under sparse sampling. By eliminating repeated calibration and explicit field reconstruction, our approach achieves high-fidelity imaging at 2.5 MHz under dynamic perturbations, establishing spectral-domain transport as a new route for robust, high-speed imaging through complex multimode channels.

# Results

## Basic concept

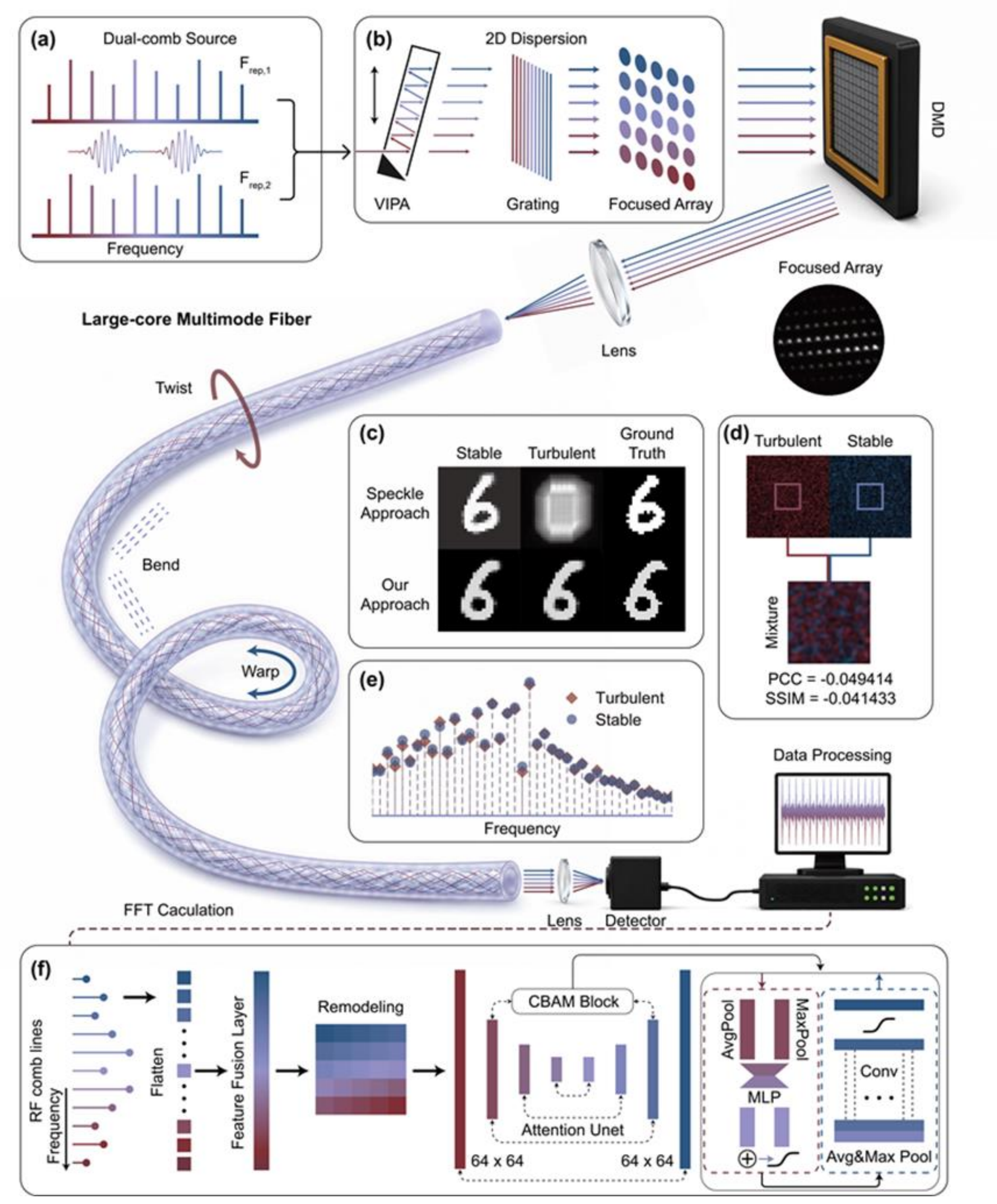

**Fig. 1 | Concept and workflow. a,** Dual-comb light source. The two combs of slight different repetition rates at $f_{r1}$ and $f_{r2}$, are used. After being reflected by the digital micromirror device (DMD) and transmitted through a 12-m-long multimode fiber, the two combs are focused onto a single-pixel detector, generating heterodyne beat notes in the radio-frequency (RF) domain. **b,** Two-dimensional spectral mapping using a virtually imaged phased array (VIPA) and a diffraction grating. **c,** Comparison of reconstruction

performance between speckle-based and spectral-encoding-based methods. **d,** Speckle pattern variations under fiber perturbations. The pearson correlation coefficient (PCC) and structural similarity index measure (SSIM) are close to zero. **e,** Measured RF comb lines under stable and perturbed fiber conditions. **f,** Dual-comb readout and neural network-based reconstruction. CBAM, convolutional block attention module; Conv, convolution; MLP, multilayer perceptron.

Figure 1 illustrates the principle of the proposed imaging system. The core idea is to encode spatial image information onto individual optical comb lines (Fig. 1(a)), establishing a one-to-one correspondence between a spatial image and a comb spectrum (Fig. 1(b)). The spectral information is then transmitted through a long-distance large-core MMF as the information carrier. Subsequently, dual-comb heterodyne detection converts the optical-frequency-domain information into the RF domain for rapid and high-resolution spectral detection. Finally, a neural network is employed to perform compressive sensing reconstruction and recover the original image.

This design offers several key advantages. First, it eliminates reliance on output speckle patterns as the primary information carrier (Fig. 1c). In conventional MMF imaging, fiber bending, twisting, and deformation modify modal phases and coupling, causing speckle calibrations to rapidly degrade as the fiber state changes (Fig. 1d). This sensitivity is evident from time-lapse measurements, where the Pearson correlation decreases rapidly under uncontrolled perturbations (Supplementary Fig. 1). By contrast, spectral encoding remains intrinsically more robust to modal perturbations during linear propagation (Fig. 1e). Second, dual-comb detection maps the encoded information to the RF domain, enabling massively parallel readout with a low-cost single-pixel photodetector and avoiding the bandwidth limitations of camera-based detection. Third, unlike conventional spectroscopic approaches, DCS requires no moving parts and enables comb-line-resolved high-resolution spectral measurements. Finally, compared with imaging schemes using optical combs and virtually imaged phased array (VIPA), the neural-network-assisted approach relaxes calibration and alignment requirements while enabling compressive image reconstruction from sparsely sampled dual-comb lines.

As shown in Fig. 1(a), a symmetric dual-comb architecture is adopted to suppress common system-induced perturbations, which affect both combs similarly and are therefore reduced relative to an asymmetric configuration [28]. Specifically, two optical combs with a small repetition-rate offset ($\Delta f << f_r$; repetition rate: $f_r$) are launched into an image-to-spectrum encoding unit (Fig. 1(b)). The encoding unit consists of a VIPA, which spectrally resolves adjacent comb lines with high dispersion, and a diffraction grating (1200 lines/mm), which separates different VIPA diffraction orders along the orthogonal direction. Together, these elements map the one-dimensional comb spectrum into a two-dimensional spot array that is focused onto a digital micromirror device (DMD). The pattern displayed on the DMD selectively redirects a portion of the spot array into the collection path, thereby encoding image information onto the amplitudes of the corresponding comb lines prior to fiber transmission (Details of experimental setup are provided in Supplementary Note 1). Following long-distance propagation, the dual combs are focused onto a photodetector, where dual-comb interferograms are generated, digitized, and Fourier transformed to compute a spectrum in the RF domain. Subsequently, the dual-comb-line peaks are extracted and arranged into a one-dimensional tensor for image reconstruction (Fig. 1(f)).

Notably, the network consists of a fully connected layer (FCL) front end followed by an attention-enhanced UNet. The one-dimensional tensor containing the comb-line information first passes through a fully connected layer for feature extraction and fusion. It is then upsampled to 4096 elements and reshaped into a $64 \times 64$ matrix. This matrix is subsequently processed by an attention-enhanced UNet incorporating the convolutional block attention module (CBAM) [38]. The UNet follows a standard architecture with four down-sampling stages and four up-sampling stages. In the skip connections, channel and spatial attention are introduced to help the model identify informative channels and spatial regions during encoder-decoder feature fusion. This design enhances informative features, suppresses irrelevant information, and improves reconstruction performance. Details of data processing and neural-network architecture are provided in Supplementary Note 2.

## Perturbation-robust image reconstruction

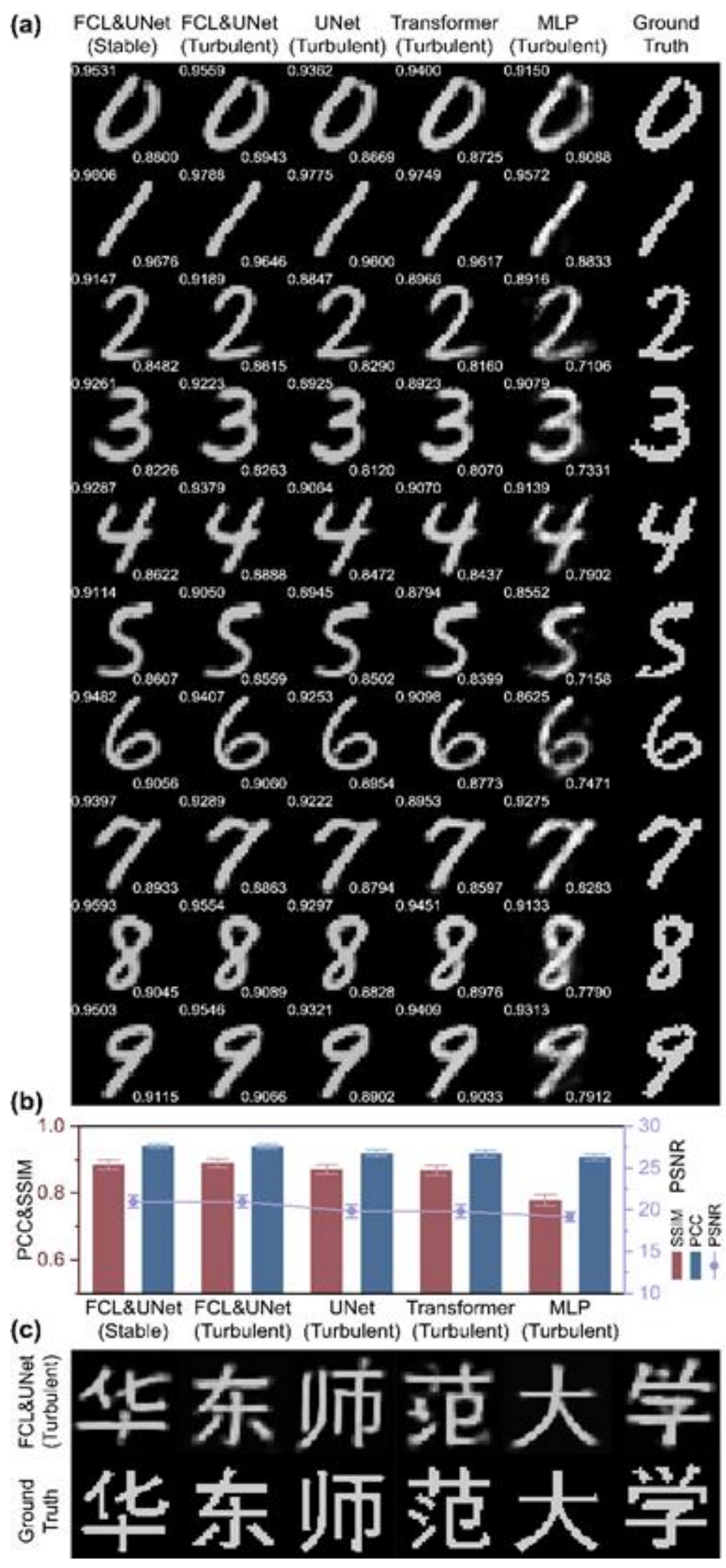

**Fig. 2 | Perturbation-robust image reconstruction under different fiber states and neural network architectures. a,** Handwritten digit reconstruction results. Columns show fully connected layer (FCL)&UNet under stable and perturbed fiber states, UNet, Transformer, and MLP under the perturbed state, and the ground truth. Overlaid values indicate PCC (upper left) and SSIM (lower right). **b,** Statistical comparison of PCC, SSIM, and peak signal-to-noise ratio (PSNR), with standard-deviation error bars indicating standard error. **c,** Compressive reconstruction of complex Chinese character targets under the perturbed fiber state.

We first assessed the robustness of spectral encoding for image reconstruction under varying fiber states. In this experiment, we employed a dual-comb source with large line spacings or repetition rates ($f_r$=25 GHz and $\Delta f$ =500 kHz). The two combs were generated using electro-optic modulators (EOMs), with one comb frequency-shifted by 40 MHz using an acousto-optic modulator (AOM) to avoid RF aliasing [23]. The reconstructed images are presented in Fig. 2.

The modified MNIST handwritten-digit results in Fig. 2(a) enable a direct comparison across both fiber conditions and neural network architectures. FCL&U-Net delivers nearly identical reconstructions in stable and perturbed states, preserving the key strokes and digit identities despite fiber deformation. The same perturbed inputs reconstructed by U-Net, Transformer, and multilayer perceptron (MLP) models likewise remain recognizable, albeit with differences in sharpness and local contrast. These results demonstrate that our approach retain the essential object information after perturbation, with reconstruction robustness originating primarily from the intrinsic stability of the spectral encoding rather than from a specific network design.

A quantitative comparison of reconstruction performance is provided in Fig. 2(b). Metrics such as Pearson correlation coefficient (PCC), structural similarity index measure (SSIM), and peak signal-to-noise ratio (PSNR) values were analyzed. Consequently, FCL&U-Net kept these values nearly unchanged before and after perturbation, while the remaining architectures also maintain high metrics under perturbed fiber states. Note that the relatively lower performance of the MLP, implemented with only two hidden layers and ReLU/sigmoid activations, arises from its limited ability to capture spatial features rather than from degradation of the spectrally encoded information.

Moreover, our approach remains effective for targets with substantially greater structural complexity. We validated this capability using information-rich images, such as Chinese characters with dense strokes and closely spaced features, while maintaining the same spectral encoding array with an effective comb-spot resolution of approximately 11×8. As shown in Fig. 2(c), the reconstructions preserve the overall character morphology and principal stroke patterns. Moderate edge broadening and background nonuniformity relative to the ground truth are still observed, primarily due to the limited spatial sampling density and the resulting constraints on sparse image reconstruction.

## Compressive reconstruction

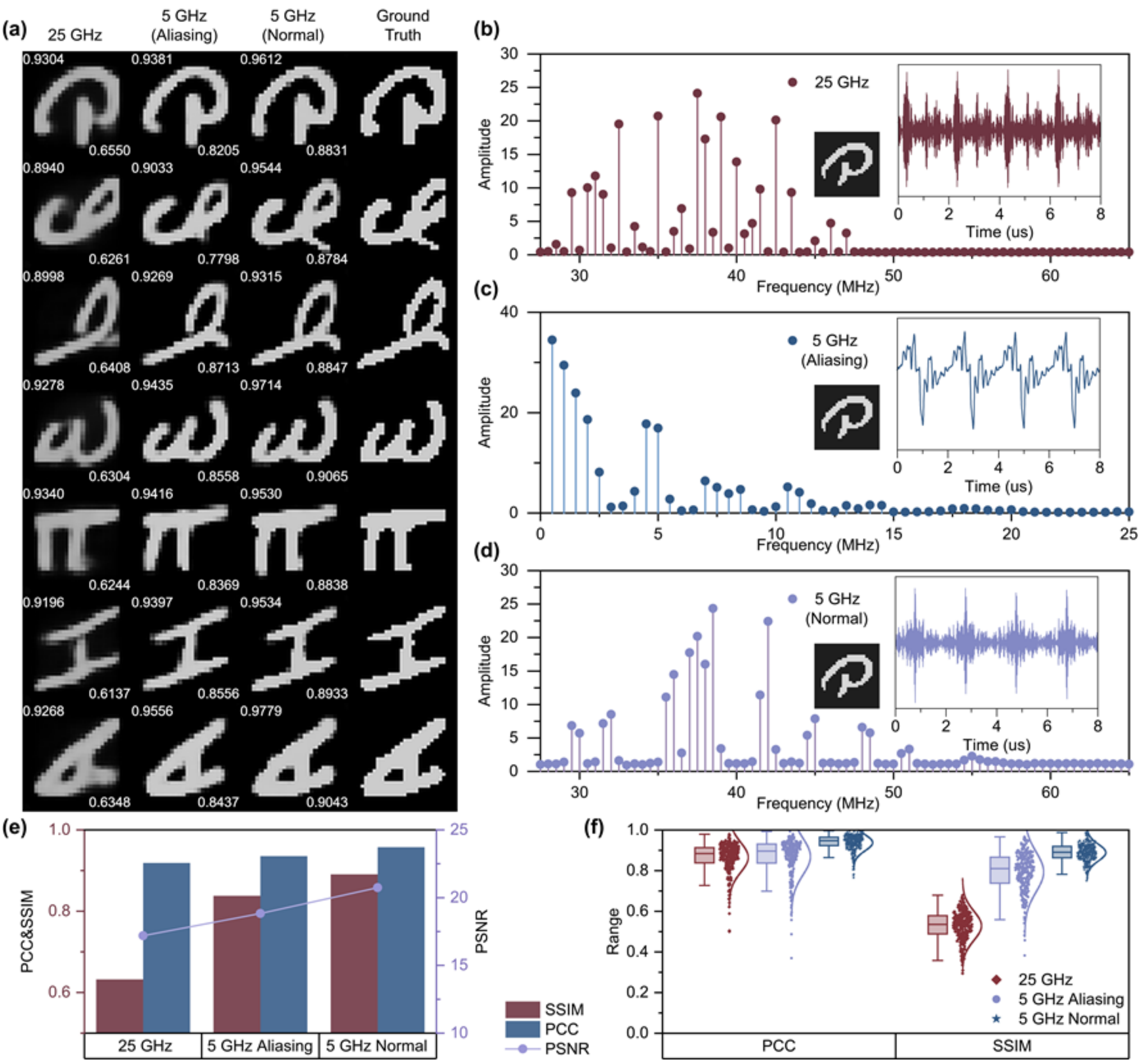

**Fig. 3 | Compressive image reconstruction. a,** Representative reconstructions obtained under different compression schemes. Spatial-domain compression was implemented with a comb repetition rate of $f_r$ = 25 GHz, whereas frequency-domain compression employed $f_r$ = 5 GHz with RF spectral aliasing. A non-aliased measurement at the same repetition rate is shown for comparison. **b–d,** Dual-comb spectra corresponding to the three measurement conditions, with the associated temporal signals and ground-truth images displayed in the right-hand panels. **e,** Mean PCC, SSIM, and PSNR values for each condition. **f,** Distributions of PCC and SSIM for 300 randomly selected image reconstructions.

We next investigated the compressive sensing capability of our method under fiber perturbations. In comb-based spectral-to-spatial mapping, each comb line represents an individual spatial sampling point, with the achievable sampling density and resolution determined by the number of comb lines and the resolving capability of the dispersive element, such as the free spectral range (FSR) of the VIPA. Furthermore, dual-comb detection must satisfy the Nyquist criterion to avoid frequency-domain aliasing, while the detector bandwidth and digitizer sampling rate limit the resolvable comb-line density and acquisition speed. These constraints hinder real-time reconstruction of high-quality images. By leveraging neural networks, we overcome this limitation and enable accurate image recovery from only a limited number of encoded channels.

We implemented compressive imaging in both the spatial and frequency domains. For spatial-domain demonstration, we set the comb parameters as $f_r$=25 GHz and $\Delta f$=500 kHz. We note two technical issues in this measurement. First, the tightly focused VIPA spot array samples only discrete spatial locations, inevitably losing information between adjacent spots. To mitigate this limitation, we slightly defocused the optical mapping system so that each spot covered a broader local region and captured additional inter-spot information. Second, we exploited the periodic dispersion of the VIPA, which maps a single comb line onto multiple spatial positions. In our system, only two to three comb lines fall within each VIPA free spectral range (FSR = 60 GHz), producing a repeated spot pattern along one spatial dimension. Consequently, at an effective spot-array resolution of approximately 9×11, identical comb lines contribute to multiple locations and generate superimposed intensity signals. Despite this non-ideal encoding, the neural network successfully disentangles the mixed comb-line information and reconstructs the images with high fidelity. The experimental results are shown in Fig. 3(a), with the corresponding dual-comb spectrum presented in Fig. 3(b). These results demonstrate that accurate image reconstruction remains possible even under spatially overlapping and mixed sampling conditions.

For the frequency-domain demonstration, we reduced $f_r$ to approximately 5 GHz, thereby increasing the number of comb lines and alleviating the limitations imposed by sparse spatial sampling. We also removed the AOM, allowing RF aliasing to occur intentionally. As shown in Fig. 3(c), the dual-comb RF spectrum becomes folded, with negative-frequency components overlapping the positive-frequency ones. Despite this spectral ambiguity, the system still reconstructs the images with high fidelity, as presented in Fig. 3(a). For comparison, Fig. 3(d) shows the corresponding unaliased dual-comb spectrum obtained with frequency shifting, together with the reconstructed images in Fig.

3(a). The similar reconstruction quality in both cases demonstrates the robustness of our approach to RF spectral aliasing and mixed frequency-domain encoding. Furthermore, Fig. 3e compares the PCC, SSIM, and PSNR values obtained for the two compressed cases with those of the uncompressed case, confirming that high-fidelity reconstruction can still be achieved under both compressed measurements, despite minor loss of fine details. In addition, Fig. 3(f) presents the reconstruction statistics for 300 randomly selected images, confirming the robustness and reproducibility of the compressive recovery across a diverse set of targets.

**High-speed imaging**

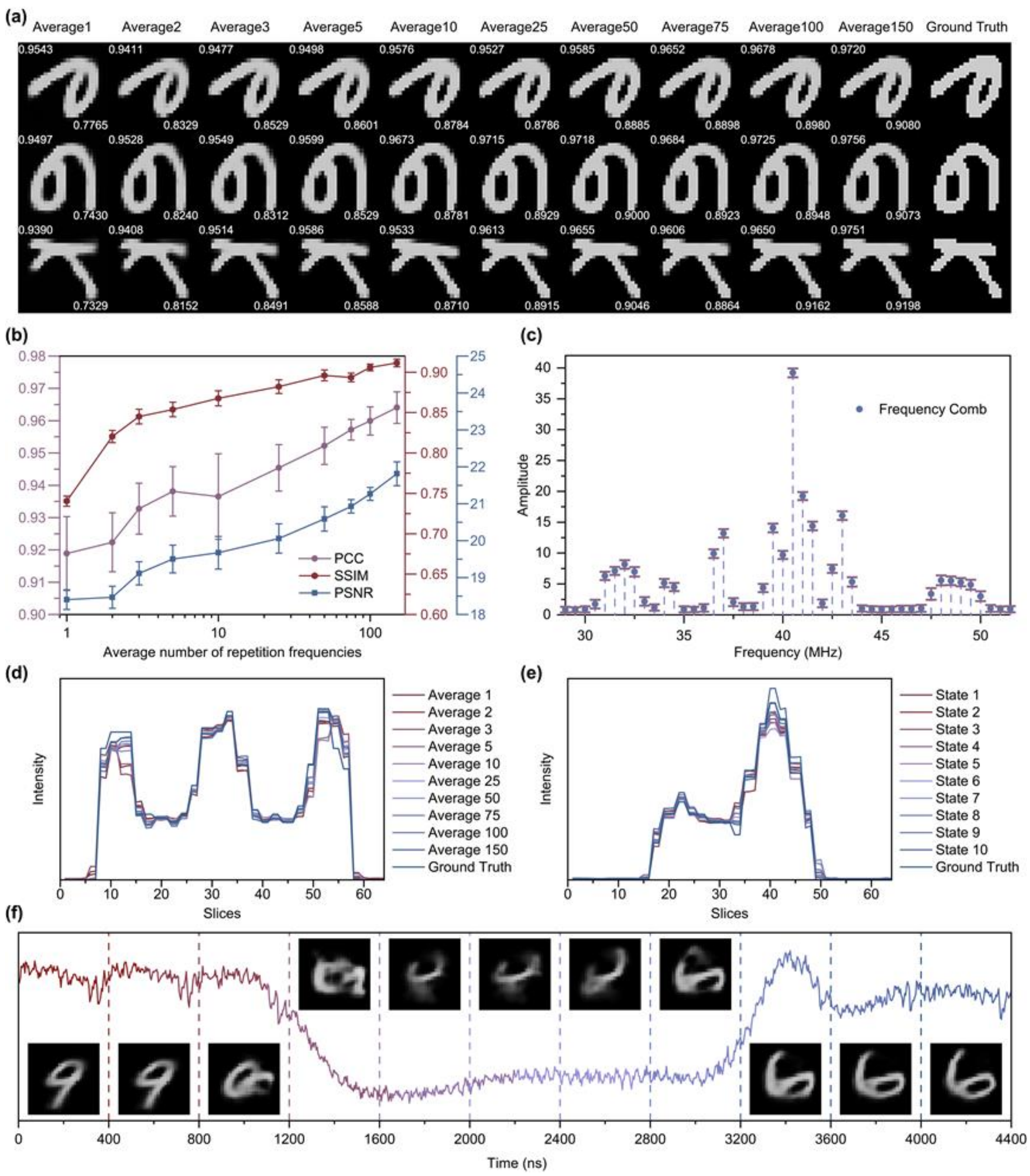

**Fig. 4 | High-speed imaging. a,** Representative reconstructed images obtained with different averaging numbers. **b,** Dependence of PCC, SSIM, and PSNR on the averaging number. **c,** Dual-comb spectral stability, with error bars denoting the standard deviation across 150 measurements. **d,** Transverse intensity profiles of the reconstructed images for different averaging conditions. **e,** Transverse intensity profiles under ten previously unseen fiber perturbation states, each recorded from a single-shot measurement. **f,** Real-time image reconstruction during DMD switching at a frame rate of 2.5 MHz. The insets show the reconstructed frames.

Finally, we extended our approach to high-speed operation. To this end, images were reconstructed at $\Delta f$, such that each frame was recovered from a single dual-comb interferogram. Representative single-shot images are shown in Fig. 4(a), with an acquisition time of only 2 μs per frame ($\Delta f$ = 500 kHz). Despite this short integration time, the reconstructions achieve a PCC above 0.9, an SSIM exceeding 0.6, and a PSNR greater than 18 dB. As expected, spectral averaging further suppresses noise and enhances image clarity and fidelity, as illustrated in Fig. 4(b).

Moreover, the system exhibits excellent reproducibility. Fig. 4(c) shows stable dual-comb readout over 150 measurements, with only minor fluctuations in the comb-line amplitudes (mean PCC to the mean spectrum = 0.995919; relative $L_2$ spectral deviation = 7.2293%). To facilitate quantitative comparison, we reduced the two-dimensional images to one-dimensional transverse intensity profiles by summing the pixel values along the vertical direction. As shown in Fig. 4(d), the profiles obtained with different averaging numbers retain the same overall shape and progressively converge to the ground-truth distribution, with an average deviation below 2%. We further evaluated ten previously unseen perturbed fiber states, and the results in Fig. 4(e) confirm the robustness and stability of the proposed approach across diverse fiber configurations.

The combination of rapid acquisition and minimal frame-to-frame variation enables the characterization of high-speed dynamics. As a demonstration, we increased the frame rate to 2.5 MHz ($\Delta f$ =2.5 MHz) and captured transient switching events of the DMD (lasting approximately 3 μs). The reconstructed image sequence in Fig. 4(f) continuously tracks the target evolution, demonstrating rapid dynamic imaging through the multimode fiber.

## Discussion

Here, we employ image-to-spectrum encoding, rather than output-speckle recovery, to overcome the challenge of perturbation-sensitive imaging through MMFs. This approach enables high-fidelity image reconstruction after fiber bending, twisting, and vibration without requiring recalibration of the fiber response. Compared with spectral image delivery through single-mode fibers, MMFs provide a larger effective mode area, thereby reducing nonlinear distortion and increasing the power tolerance for pulsed transmission [4–6]. Examples of the spectral distortion experienced by the combs after propagation through a piece of single-mode fiber are provided in Supplementary Fig. 2.

In our system, image-to-spectrum encoding is implemented using a VIPA, which has previously been employed in time-encoded imaging [39], Brillouin and biomechanical microscopy [40,41], and spectral-mode demultiplexing [42]. Here, we extend this technique to flexible MMF imaging and further incorporate compressive recovery. Combined with neural-network-based decoding, images can be reconstructed from sparse, noisy, or aliased dual-comb measurements [15–18]. Recently, nonlinear up-conversion has also been explored for robust MMF image transport; however, it requires high pump power and is subject to phase-matching constraints [19]. In contrast, our method is a linear approach that maintains high reconstruction fidelity while requiring less optical power without phase-matching limitations. Moreover, by leveraging dual-comb technology, our approach enables high-speed image readout with a single-pixel detector.

Currently, our system is limited by the number of comb lines and the detector bandwidth. Increasing the number of comb lines and extending the spectral coverage would improve image quality, while higher-bandwidth balanced detectors and faster digitizers would enlarge the beat-note window and alleviate the trade-off between frame rate and reconstruction fidelity. Joint training with mixed datasets, or pretraining followed by task-specific fine-tuning, could further enhance generalizability [43]. Miniaturization and integration are next steps. On-chip comb sources [21,44], compact dispersers [42], integrated gratings or metasurfaces [45,46], and fiber-tip micro-optics [45] could enable a more practical and miniaturized platform for fiber-probe imaging systems.

Nevertheless, our approach with the unique combination of several technical features enables a broad range of applications. Narrow optical access and flexible light delivery are attractive for minimally invasive fiber endoscopy [1] and single-fiber imaging [47]. Perturbation-tolerant image transport enables distal imaging, remote sensing, and operation under complex enviroments [45]. High-speed single-pixel readout supports high-throughput imaging flow cytometry [48] and real-time optical monitoring [11, 49]. The large mode area and high power tolerance of MMFs [50] facilitate long-distance information transport [3]. In addition, the compatibility with compact spectral encoding and dual-comb technologies makes the platform promising for fiber-probe Brillouin and biomechanical imaging [41] and future miniaturized fiber-based photonic instruments.

## Acknowledgements

This work was supported by Quantum Science and Technology-National Science and Technology Major Project (2023ZD0301000), Chongqing Special Major Project for Technological Innovation and Application Development (CSTB2025TIAD-STX0024), and Shanghai Municipal Science and Technology Major Project (2019SHZDZX01).

## Author contributions

M.Y., M.L., and Z.W. conceived the idea and designed the experiments. M.L. conducted the experiment and drafted the manuscript. B.S., Z.R. W., X.M. and X.Z. contributed to the optical setup and system optimization. Y.C. and Q.W. build the comb source. Z.W. contributed to data interpretation. Z.W., M.Y., and H.Z. revised the manuscript. All authors provided comments and suggestions for improvements.

## Competing interests

The authors declare no competing interests.

## Data availability

The data that support the findings of this study are available from the corresponding author upon reasonable request.

## Author information

Correspondence and requests for materials should be addressed to M.Y. (myan@lps.ecnu.edu.cn).